\documentclass[a4paper,fleqn,usenatbib]{mnras}

\usepackage[T1]{fontenc}
\usepackage{ae,aecompl}

\usepackage{graphicx}	
\usepackage{amsmath}	
\usepackage{amssymb}	

\title[Spin-up and spin-down of long period X-ray pulsars]{Understanding the coexistence of spin-up and spin-down behaviors in long period X-ray pulsars}

\author[W. Wang \& H. Tong]{
W. Wang,$^{1,2}$\thanks{E-mail: wangwei2017@whu.edu.cn}
H. Tong$^{3}$\thanks{E-mail: htong\_2005@163.com}
\\
$^{1}$School of Physics and Technology, Wuhan University, Wuhan 430072, China \\
$^{2}$WHU-NAOC Joint Center for Astronomy, Wuhan University, Wuhan 430072, China \\
$^{3}$School of Physics and Electronic Engineering, Guangzhou University, Guangzhou 510006, China
}

\date{Accepted XXX. Received YYY; in original form ZZZ}

\pubyear{2015}

\begin{document}
\label{firstpage}
\pagerange{\pageref{firstpage}--\pageref{lastpage}}
\maketitle

\begin{abstract}
Assuming the wind-fed accretion magnetars in long period X-ray pulsars, we calculated the rotational evolution of the neutron stars. Our calculations considered the effects of the magnetic field decay in magnetars. The results show that wind-fed accretion magnetars can evolve to the long period X-ray pulsars with a spin period much longer than 1000 s. The spin-down trend observed in 4U 2206+54 like sources is expected when the young X-ray binary systems are on the way to their equilibrium period. Detailed calculations showed that its spin-down may be affected by accretion with outflows or accretion while spin-down. Due to the magnetic field decay in magnetars, wind-fed accretion magnetars will have a decreasing equilibrium period for a constant mass accretion rate. For 2S 0114+65, the spin-up rate due to magnetic field decay is one order of magnitude smaller than the observations. The spin-up rate of 2S 0114+65 may be attributed to the formation of a transient disk during wind accretion. The slowest X-ray pulsar AX J1910.7+0917 would be a link source between 4U 2206+54 and 2S 0114+65.
\end{abstract}

\begin{keywords}
stars: magnetars -- stars: neutron -- X-ray: binaries -- pulsars: individual (4U 2206+54, 2S 0114+65, AX J1910.7+0917)
\end{keywords}



\section{Introduction}

Accretion-powered pulsars are neutron star systems by accreting materials from the companion stars. The spin period evolution of the accretion-powered pulsar depends on the accretion form, i.e. disk-fed accretion or wind-fed accretion, as well as the neutron star magnetic field. Thus the study of spin period evolution is an important approach on the neutron star magnetic field and accreting materials on neutron stars (Bildsten et al. 1997; Frank et al. 2002). Long period X-ray pulsars are a special kind of accretion-powered pulsars in high mass X-ray binaries (Walter et al. 2015), in which the neutron stars generally have rotational period longer than 1000 seconds.

There are some typical long period X-ray pulsars in observations at present: 4U 2206+54 ($P_{\rm s} \sim 5560 \ \rm s$, Reig et al. 2009; Wang 2009), 2S 0114+65 ($P_{\rm s} \sim 10^4 \ \rm s$, Finley et al. 1992; Corbet et al. 1999), AX J1910.7+0917 ($P_{\rm s} \sim 3.6\times 10^4 \ \rm s$, Sidoli et al. 2017), SXP1062 ($P_{\rm s} \sim 1062 \ \rm s$, Henault-Brunet et al. 2012; Haberl et al. 2012), and SXP 1323 ($P_{\rm s} \sim 1323 \ \rm s$, Carpano et al. 2017). Though the number of long period X-ray pulsars is limited, the physical origin of the long rotation period and properties of these accreting systems are involved in much interest. The physical mechanism of the neutron star's long rotational period is still an open question (Li \& van der Heuvel 1999; Ikhsanov 2007; Shakura et al. 2012). In addition, these long period X-ray pulsars undergo the fast and long-term spin evolution. The neutron star in 4U 2206+54 shows a long-term spin-down trend with a mean rate of $\sim 5\times 10^{-7}$ s s$^{-1}$ in last 15 years (Wang 2013; Torrejon et al. 2018). While the neutron star in 2S 0114+65 undergoes a long-term spin-up trend with a mean rate of $\sim - 10^{-6}$ s s$^{-1}$ in last 30 years (Hall et al. 2000; Wang 2011; Sanjurjo-Ferrrin et al. 2017), though sometimes it shows a short term variation of spin-down and spin-up behaviours (Hu et al. 2017). The period derivatives of these long period X-ray pulsars are not explored in full details yet. Especially, why do some X-ray pulsars (e.g., 4U 2206+54) show the long-term spin-down trend, while some (e.g., 2S 0114+65) show the long-term spin-up trend? This issue is still quite uncertain.

Previously, Finger et al. (2010) discussed the long pulsation period of 4U 2206+54 in the subsonic propeller model (Ikhsanov 2007). They estimated the magnetic field using the maximum spin-down torque for the spin-down rate. Reig et al. (2012) discussed the long pulsation period of 4U 2206+54 in the quasi-spherical accretion model (Sakura et al. 2012). Wang (2013) discussed possible evolution between 4U 2206+54 and 2S 0114+65. Torrejon et al. (2018) confirmed the long term spin-down behavior of 4U 2206+54. Neglecting the spin-up torque, they obtained lower limits on the magnetic field of 4U 2206+54 in the quasi-spherical settling accretion scenario (Torrejon et al. 2018).

Hall et al. (2000) first reported the spin-up rate for 2S 0114+65, and discussed the possible origin of the long pulsation period according to the model proposed by Li \& van den Heuvel (1999). Bonning \& Falanga et al. (2005), Farrel et al. (2008), Wang (2011) confirmed the long term spin-up rate. Farrel et al. (2008) pointed that the spin-up rate of 2S 0114+65 may be due to accretion, giving a mass accretion rate about $10^{-8} \rm M_{\odot} yr^{-1}$, which is too high for the X-ray luminosity. Sanjurjo-Ferrrin et al. (2017) further confirmed the long term spin-up rate for 2S 0114+65, and discussed the long pulsation period in terms of the quasi-spherical settling accretion (Sakura et al. 2012), but the spin-up behavior was not discussed. Hu et al. (2017) proposed possible transient accretion disk may account for the spin-up of 2S 0114+65. No quantitative calculation is done for the the possible transient disk in Hu et al. (2017).

Based on the present observations of 4U 2206+54 and 2S 0114+65, it is found that previous observational discussion and theoretical modeling are far from complete. We try to model both the long pulsation period and period derivative simultaneously. Furthermore, from the long term rotational evolution of wind accreting magnetars, there may be evolutional links between the long period X-ray pulsars (Wang 2013; Tong \& Wang 2019). We try to understand the rotational behavior of long period X-ray pulsars from the wind accreting magnetar point of view. The strong magnetic field
may account for the neutron star's long rotational period (Li \& van der Heuvel 1999; Reig et al. 2012). Considering the magnetic field decay of magnetars, the accreting magnetars may experience both spin-down and spin-up during its evolution (Urpin et al. 1998; Popov \& Turolla 2012; Fu \& Li 2012). Previous work considered magnetic field decay for accreting neutron stars with normal magnetic field (Urpin et al. 1998), or the accreting magnetar candidate SXP 1062 (Popov \& Turolla 2012; Fu \& Li 2012). Compared with Popov \& Turolla (2012) or Fu \& Li (2012), we considered the modification of accretion torque in the wind accretion case and effect of transient disks during wind accretion.

The theoretical basis to calculate the rotational evolution of wind accreting magnetars are induced in Section 2, including the magnetic field evolution, and wind-fed accretion torques. Then the long-term spin-down trend of 4U 2206+54 and the long-term spin-up trend of 2S 0114+65 are modeled quantitatively in section 3. The discussion and conclusion are presented in Section 4.

\section{Rotational evolution of wind accreting magnetars}

\subsection{Magnetic field evolution of magnetars}

The magnetic field of the normal pulsars ($B\sim 10^{12}$ G) is expected to decay on timescale about $10^6$ years (Shapiro \& Teukolsky 1983). In the neutron star low mass X-ray binaries, the accreted material will also affect the evolution of the neutron star's magnetic field (Psaltis 2006). Due to the short life time of a massive star and the low mass accretion rate, accretion induced magnetic field decay in high mass X-ray binaries may be negligible (Psaltis 2006).

Magnetars are assumed to be young neutron stars whose emissions are powered by the magnetic field energy, i.e., due to the decay of the magnetic field.
The magnetic field decay in magnetars is coupled with its thermal evolution (Vigano et al. 2013).
Simplified treatment of magnetic field decay showed that the magnetic field may evolve with
time in the following analytical form (Aguilera et al. 2008; Popov \& Turolla 2012)
\begin{equation}\label{eqn_B_decay_1}
B(t) = \frac{B_0 \exp{(-t/\tau_{\rm O})}}{1+ \tau_{\rm O}/\tau_{\rm H} (1-\exp{(-t/\tau_{\rm O})})} +B_{\rm fin},
\end{equation}
where $B_0$ is the initial magnetic field, $\tau_{\rm O}$ and $\tau_{\rm H}$ are the Ohmic and Hall decay timescale, respectively,
$B_{\rm fin}$ is the final relic magnetic field.
Considering that the core field and crustal field may evolve and contribute differently (Heyl \& Kulkarni 1998; Geppert et al. 2000), a power law form of magnetic field decay may be employed (Colpi et al. 2000; Dall'Osso et al. 2012)
\begin{equation}\label{eqn_B_decay_2}
\frac{d B}{d t} = -a B^{1+\alpha},
\end{equation}
where $a$ and $\alpha$ are the model parameters. For different avenues of magnetic field decay, the model parameters are different (Colpi et al. 2000): $\alpha=5/4$ for ambipolar diffusion, $\alpha=1$ for crustal Hall cascade.

\begin{figure}
  \includegraphics[width=0.45\textwidth]{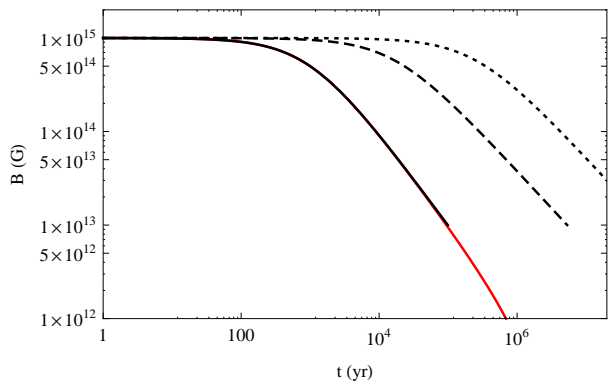}
  \caption{Magnetic field decay in the case of magnetars. The red solid line is the result of equation (\ref{eqn_B_decay_1}). The black lines are the results of equation (\ref{eqn_B_decay_2}): the dotted ($a$=0.01, $\alpha=5/4$ for ambipolar diffusion in the irrotational mode), dashed ($a$=0.15, $\alpha$=5/4 for ambipolar diffusion in solenoidal mode), and solid lines ($a$=10, $\alpha$=1 for crustal Hall cascade) are for different avenues of magnetic field decay (Colpi et al. 2000). The red solid line and the black solid line coincide with each other.}
  \label{fig_B_decay}
\end{figure}

Figure \ref{fig_B_decay} shows the decay of magnetic field for typical magnetar parameters.
The red solid line shows the magnetic field decay, for an initial magnetic field of $10^{15} \ \rm G$, with Ohmic and Hall decay
time scale of $10^6$ yr and $10^3$ yr, respectively. The final relic magnetic field $B_{\rm fin}$ is assumed to be
similar to that of millisecond pulsars, $10^8 \ \rm G$. It can be taken as zero in the present case. The black dotted,
dashed, and solid lines are the magnetic field decay described by the power law form. Different patterns are for different
avenues of magnetic field decay, core or crustal etc (see Colpi et al. 2000 for details). The red solid line and the black
solid line coincide with each other. Therefore, equation (\ref{eqn_B_decay_1}) and (\ref{eqn_B_decay_2}) gives similar results
considering Hall decay or the crustal field.  Equation (\ref{eqn_B_decay_1}) will be used in the following.

Recently, there are increasing candidates of accreting magnetars proposed, e.g. ultra-luminous X-ray pulsars (Bachetti et al. 2014; Israel et al. 2017; Bachetti et al. 2019), long period X-ray pulsars (Reig et al. 2012; Sanjurjo-Ferrrin et al. 2017; Tong \& Wang 2019). In the case of ultraluminous X-ray pulsars, the mass accretion rate may be highly super-Eddington. A huge amount of accreted matter may result in the decay of the magnetar's magnetic field (Zhang \& Kojima 2006; Pan et al. 2016). Long period X-ray pulsars may be wind accreting magnetars with mass accretion rate about $10^{15}$-$10^{16} \ \rm g \ s^{-1}$. For a typical life time of the high mass X-ray binary of $\sim 10^6$ years, the total mass of accreted matter is about $10^{-5} \ \rm M_{\odot}$. According to the study of accreting low mass X-ray binaries, this amount of accreted matter will not affect the stellar magnetic field significantly (Shibakazi et al. 1989; Zhang \& Kojima 2006). Therefore, the magnetic field decay in wind accreting magnetars may be solely determined by the physics inside the magnetar (as described above).

\subsection{Equilibrium period}

In a high mass X-ray binary, the neutron star may capture the stellar wind material and
manifest itself as a X-ray pulsar. The light cylinder radius of a pulsar is defined as the radius when the corotational velocity is equal to the speed of light,
\begin{equation}
R_{l} = \frac{P c}{2\pi}.
\end{equation}
The light cylinder radius may be taken as the boundary of the neutron star's magnetosphere. Inside the magnetospheric radius, the material will be controlled by the magnetic field and try to corotate with the neutron star.
Besides the gravitational pull of the central neutron star, the material will also feel the centrifugal force.
If the centrifugal force is larger than the gravitational pull, the accreted matter still can not fall onto
the neutron star. It may even be thrown away by the central neutron star. This is the so-called ``propeller effect" (Illarionov \& Sunyaev 1975). The criterion for the propeller phase is that the magnetspheric radius is larger than the neutron star corotation radius
\begin{equation}
R_{\rm m} > R_{\rm co},
\end{equation}
where $R_{\rm co}\equiv (G M_{\rm ns}/4\pi^2)^{1/3} P^{2/3}$ is the corotation radius. It is defined as the radius where the Keplerian angular velocity is equal to the neutron star angular velocity. The magnetospheric radius is defined as the radius when the magnetic energy density is equal to the kinetic energy density of the accreted matter (Shapiro \& Teukolsky 1983),
\begin{equation}\label{eqn_Rm}
R_{\rm m} = \left( \frac{\mu^4}{2 G M_{\rm ns} \dot{M}_{\rm acc}^2} \right)^{1/7},
\end{equation}
where $\mu$ is the magnetic dipole moment of the neutron star\footnote{The magnetic dipole moment is related to the surface magnetic field as (Shapiro \& Teukolsky 1983; Lyne \& Graham-Smith 2012): $\mu =1/2 B_{\rm p} R^3 = B_{\rm s} R^3$, where $B_{\rm p}$ and $B_{\rm s}$ are the surface magnetic field at the magnetic pole and equator, respectively, $R$ is the neutron star radius. }, $\dot{M}_{\rm acc}$ is the mass accretion rate at the magnetospheric boundary. In the propeller phase, the mass accretion rate that can fall onto the neutron star is much smaller than $\dot{M}_{\rm acc}$. Only in the accretion phase, $\dot{M}_{\rm acc}$ is the mass accretion rate onto the neutron star. For disk accretion case, there will be a factor of $0.5$ in equation (\ref{eqn_Rm}) (Shapiro \& Teukolsky 1983). Since we are mainly considering wind accreting magnetars, this $0.5$ factor is not included in equation (\ref{eqn_Rm}). In the presence of transient accretion disks, omission of this factor will only result in quantitative changes.

When the magnetospheric radius is smaller than the corotation radius, the accreted material can fall onto the neutron star surface. This is the so-called ``accretion phase''. Independent of the various proposals of torques during the propeller or accretion phase (Ghosh \& Lamb 1979; Illarionov \& Sunyaev 1975; Bozzo et al. 2008; Popov \& Turolla 2012; Fu \& Li 2012), an equilibrium period may often be assumed for an accreting neutron star (Stella et al. 1986; Walters \& van Kerkwijk 1989; Ho et al. 2014). The equilibrium period is defined as when the magnetospheric radius is equal to the corotation radius,
\begin{equation}\label{eqn_Peq}
P_{\rm eq} = 6.7\times 10^3 \mu_{33}^{6/7} \dot{M}_{\rm acc,15}^{-3/7} \ \rm s.
\end{equation}
From the expression of the equilibrium period, a qualitatively picture of the rotational evolution of wind accreting magnetar
can be obtained:
\begin{itemize}
  \item Early in the evolution stage, the magneatar will spin down (due to pulsar torque or propeller torque) until
  it reaches the equilibrium period.
  \item Later, the magnetar's magnetic field will start to decay. The corresponding equilibrium period
  will also decrease (for a constant mass accretion rate). The accreting magnetar will spin-up and try to catch the new equilibrium period.
\end{itemize}
The existence of both spin-down and spin-up of wind accreting magnetars are quantitatively consistent with the observations
of long period X-ray pulsars.

\subsection{Rotational evolution of wind accreting magnetars}

The following calculations are done in analog with those of Popov \& Turolla (2012), Fu \& Li (2012). During
the ejector phase, the magnetar is spun down by magnetospheric torque (Spitkovsky 2006; Kou \& Tong 2015).
However, for the purpose here, the rough magnetic dipole radiation is enough.
The corresponding period derivative in this stage is
\begin{equation}\label{eqn_Pdot_ejector}
\dot{P} = \frac{8\pi^2 \mu^2}{3 I c^3 P}.
\end{equation}

During the propeller phase, the spin-down torque of the wind accreting magnetar can be done similar to that
of disk accreting systems (Ghosh \& Lamb 1979; Ho et al. 2014). This is equivalent to assuming a maximum efficient torque (Francischelli \& Wijers 2002; Popov \& Turolla 2012). The corresponding spin down torque is
\begin{equation}\label{eqn_Npropeller_1}
N_{\rm tot} = \dot{M}_{\rm acc} R_{\rm m}^2 \Omega_{\rm K}(R_{\rm m}) \left( 1- \frac{\Omega}{\Omega_{\rm K}(R_{\rm m})}  \right).
\end{equation}
The factor $1-\Omega/\Omega_{\rm K}(R_{\rm m})$ ensures that the accreting neutron star can reach accretion equilibrium in the long run. It unites the spin up torque in the accretion phase and the spin down torque in the propeller phase. For accretion neutron stars near the equilibrium, the torque behavior may be more complicated than this (Ghosh \& Lamb 1979; Ho et al. 2014).
In the case of wind accretion, the specific angular momentum carried by the accretion material is smaller than the Keplerian value. Therefore,
an efficiency parameter $\zeta$ may be introduced in the torque (Urpin et al. 1998; Ikhsanov 2007)
\begin{equation}\label{eqn_Npropeller_2}
N_{\rm tot} = \zeta \dot{M}_{\rm acc} R_{\rm m}^2 \Omega_{\rm K}(R_{\rm m}) \left( 1- \frac{\Omega}{\Omega_{\rm K}(R_{\rm m})}  \right).
\end{equation}
The value of $\zeta$ may ranges from 0.01 to 0.1 for the wind accretion case. In equation (\ref{eqn_Npropeller_2}), both the spin-down torque
in the propeller phase and spin-up torque in the accretion phase will be decreased by the $\zeta$ factor. The duration of the propeller phase will be longer due to decrease of the spin-down torque. While, the equilibrium period will be the same as equation (\ref{eqn_Npropeller_1}) provided that the magnetic field is the same. The corresponding period derivative in this stage is
\begin{equation}\label{eqn_Pdot_propeller}
\dot{P} = - \zeta \dot{M}_{\rm acc} R_{\rm m}^2 [ \Omega_{\rm K}(R_{\rm m})- 2\pi/P ] \ \frac{P^2}{2\pi I}.
\end{equation}

\begin{figure}
  \includegraphics[width=0.45\textwidth]{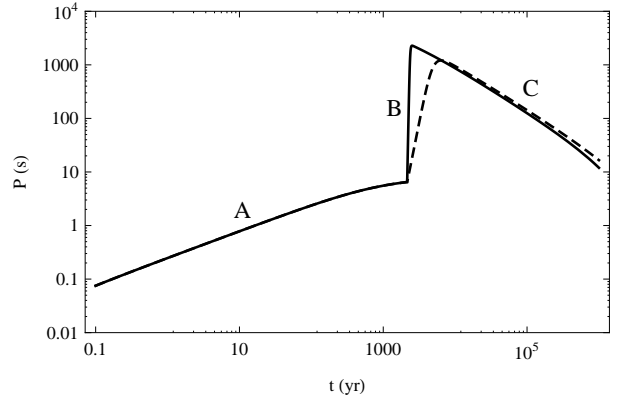}
  \caption{Rotational evolution of a typical wind accreting magnetar ($B_0 = 10^{15} \ \rm G$ and $\dot{M}_{\rm acc} = 10^{15} \ \rm g \ s^{-1}$). The solid and dashed lines are for $\zeta=1, \ 0.1$, respectively. The early spin-down is due to spin down torque in the ejector (segment A) or propeller phase (segment B). The later spin-up (segment C) is due to a decaying magnetic field, which results in a decreasing equilibrium period. For segment A, the rotational evolution of the neutron star is governed by equation (\ref{eqn_Pdot_ejector}). For segment B and C, the rotational evolution is determined by equation (\ref{eqn_Npropeller_2}). During the calculations, the rotational evolution is solved simultaneously with the magnetic field evolution.}
  \label{fig_P_evo}
\end{figure}

Equation (\ref{eqn_Npropeller_2}) is the similar to that employed by Popov \& Turolla (2012) when studying the rotational evolution of the accreting magnetar candidate SXP 1062. In addition, it also considers the decrease of specific angular moment in the case of wind accretion (Urpin et al. 1998; Ikhsanov 2007). In the case of a transient disk, the $\zeta$ parameter reflects the duty cycle of the disk. The conclusion will be the same by employing other torque formulae (e.g., Fu \& Li 2012; Ikhsanov 2007, see the Appendix for discussions). The result is mainly determined by combination of the two ideas:  equilibrium period and magnetic field decay. Typical calculation is shown in Figure \ref{fig_P_evo}.
An initial magnetic field of $B=10^{15} \ \rm G$ (surface magnetic field at the equator), and mass accretion rate of $\dot{M}= 10^{15} \ \rm g \ s^{-1}$ are assumed. During the calculations, the rotational evolution is solved simultaneously with the magnetic field evolution. The evolution of the magnetic field is described by Equation (\ref{eqn_B_decay_1}). Consistent with the qualitatively analysis, the rotation period of the accreting magnetar will first increase, and then decreases in the later state. The rotational period can be much higher than $1000$ seconds.

There may be uncertainties in modeling the torque in the accretion phase. Using the general idea of accreting equilibrium,
the rotational evolution in the accretion phase can be modeled by the expression of the equilibrium period (equation \ref{eqn_Peq}).
The time derivative of the equilibrium period is
\begin{equation}\label{eqn_Peq_evo}
\dot{P}_{\rm eq} = \frac67 \frac{P_{\rm eq}}{\mu/\dot{\mu}} -\frac37 \frac{P_{\rm eq}}{\dot{M}_{\rm acc}/(d\dot{M}_{\rm acc}/dt)}.
\end{equation}
For a constant mass accretion rate, the evolution of the equilibrium period is determined by the evolution of the
magnetic field. A decreasing magnetic field will result in a decreasing equilibrium period, and vice versa.
Defining the magnetic field evolution time scale $\tau_{\rm B} =B/\dot{B} = \mu /\dot{\mu}$, the first term in
equation (\ref{eqn_Peq_evo}) is:
\begin{equation}\label{eqn_Pdot_B_evo}
\dot{P}_{\rm eq} = \frac67 P_{\rm eq}/ \tau_{B}.
\end{equation}
For segment C in Figure \ref{fig_P_evo}, our calculations showed that the result is the same by adopting Equation (\ref{eqn_Npropeller_2}) (coupled with the magnetic field evolution) or Equation (\ref{eqn_Pdot_B_evo}). Thus, in the following calculations, Equation (\ref{eqn_Npropeller_2}) will be used.

\section{Comparison with the observations of long period X-ray pulsars}

\subsection{Explanation of the spin-down of 4U 2206+54}

The X-ray luminosity of 4U 2206+54 is about $10^{35} \ \rm erg \ s^{-1}$. It is spinning down with a rate about $5\times 10^{-7} \ \rm s \ s^{-1}$ in the last 15 years (Finger et al. 2010; Reig et al. 2012; Wang 2013; Torrejon et al. 2018). The corresponding mass accretion rate that can fall onto the neutron star is about $10^{15} \ \rm g \ s^{-1}$. If it is in the propeller phase, then the mass accretion rate at the magnetospheric radius will be even larger. Previous studies showed that only 2 to 3 percent of the total mass accretion rate can fall onto the neutron star during the propeller phase (Cui 1997; Zhang et al. 1998; Tsygankov et al. 2016). Then the total mass accretion rate will be about $50 \times 10^{15} \ \rm g \ s^{-1}$. If the neutron star is still in the propeller regime, then it should have a very high magnetic field $>5.6\times 10^{15} \ \rm G$, in order to result in a magnetospheric radius larger than the corotational radius. This could be possible. But the probability that we can observe such high magnetic field neutron stars may be small (from the distribution of neutron star magnetic field, Popov et al. 2010).

The following two scenarios may be more probable than the above case:
\begin{itemize}
  \item Accretion with outflows. For 4U 2206+54, the mass accretion rate that can reach the neutron star surface is about $\dot{M}_{\rm acc} \sim 10^{15} \ \rm g \ s^{-1}$. At the same time, there may be an outflow due to Compton heating that carry away the angular momentum and result in the spin-down of the central neutron star (Illarionov \& Kompaneets 1990). This scenario has also been discussed by Dai et al. (2006), Ertan et al. (2007) for other sources. The mass outflow rate may be only a fraction of the mass accretion rate. At the same time, the spin-up torque due to accretion onto the neutron star will also be suppressed in the wind accretion case. Therefore, both the spin-up and spin-down torque will be suppressed compared with equation (\ref{eqn_Npropeller_1}). Then the total torque may have the form of equation (\ref{eqn_Npropeller_2}).
  \item Accretion while spin-down. Similar to the disk accretion case (Ghosh \& Lamb 1979; Wang 1995), the neutron star may spin down in the accretion phase due to the coupling of magnetic field and accretion flow. Quantitatively, a critical fastness can be introduced in the total torque. For example, in equation (\ref{eqn_Npropeller_2}), the modified torque will be: $N_{\rm prop} \propto 1-(1/\omega_{\rm c}) \times \Omega/\Omega_{\rm K} (R_{\rm m})$. Here $\omega_{\rm c} \approx 0.35$ is the critical fastness (Ghaosh 1995). For this value of critical fastness, when $0.5 R_{\rm co} < R_{\rm m} < R_{\rm co}$, the neutron star will spin down during accretion phase. In the presence of a critical fastness, the equilibrium period is larger by a factor of $1/\omega_{\rm c}$. While the trend of the long term rotational evolution (in Figure \ref{fig_P_evo}) is unchanged.
\end{itemize}

\begin{figure}
\begin{minipage}{0.45\textwidth}
 \includegraphics[width=0.95\textwidth]{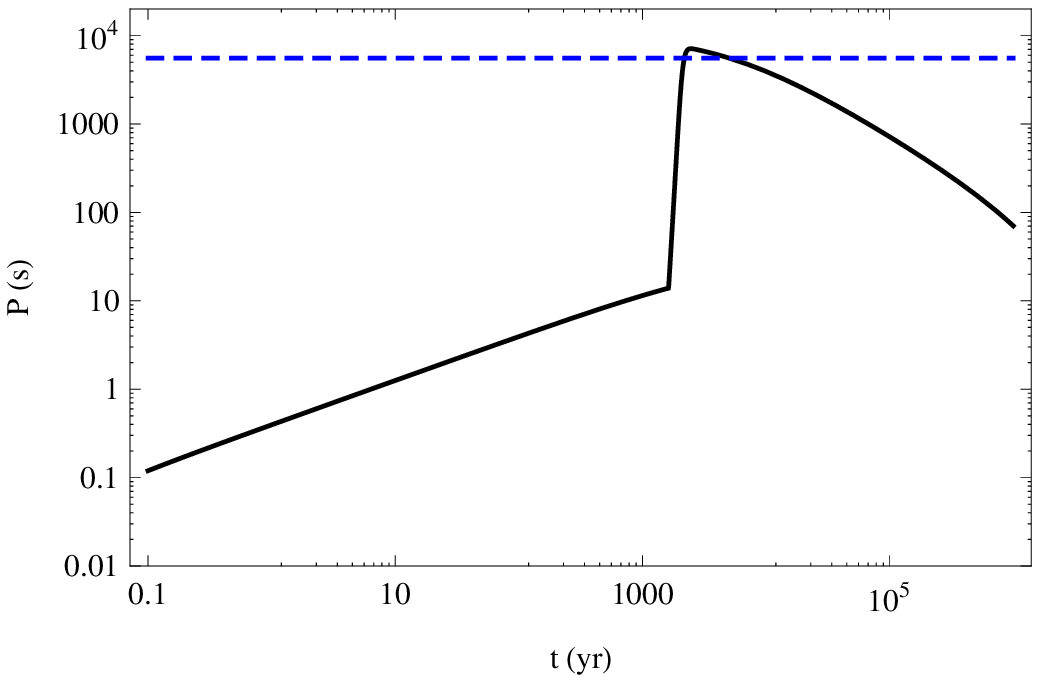}
\end{minipage}
\begin{minipage}{0.45\textwidth}
 \includegraphics[width=0.95\textwidth]{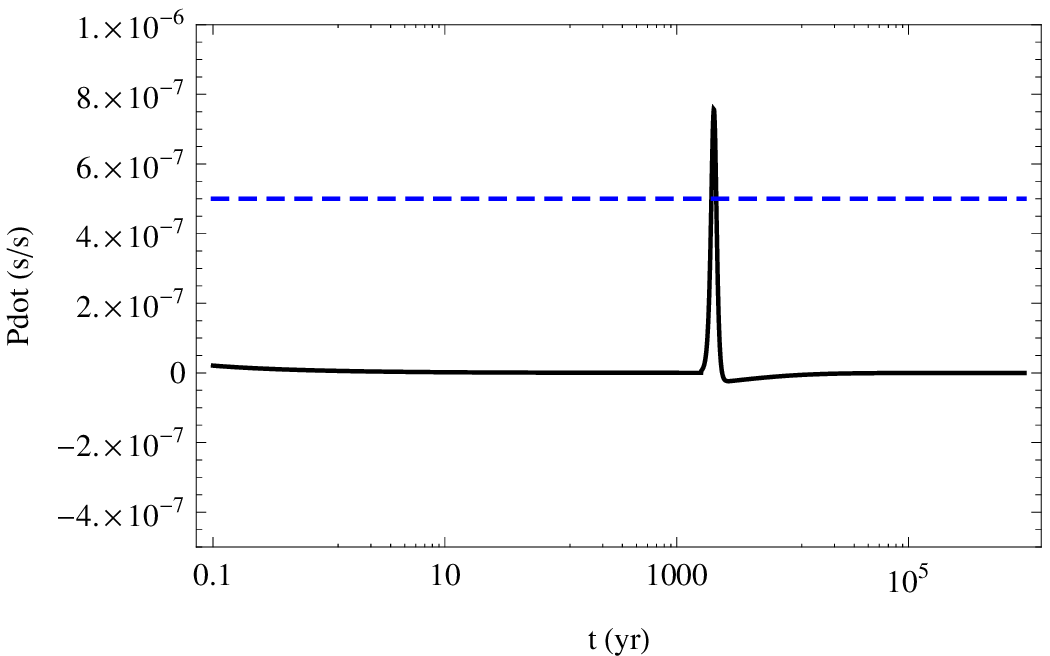}
\end{minipage}
\begin{minipage}{0.45\textwidth}
 \includegraphics[width=0.95\textwidth]{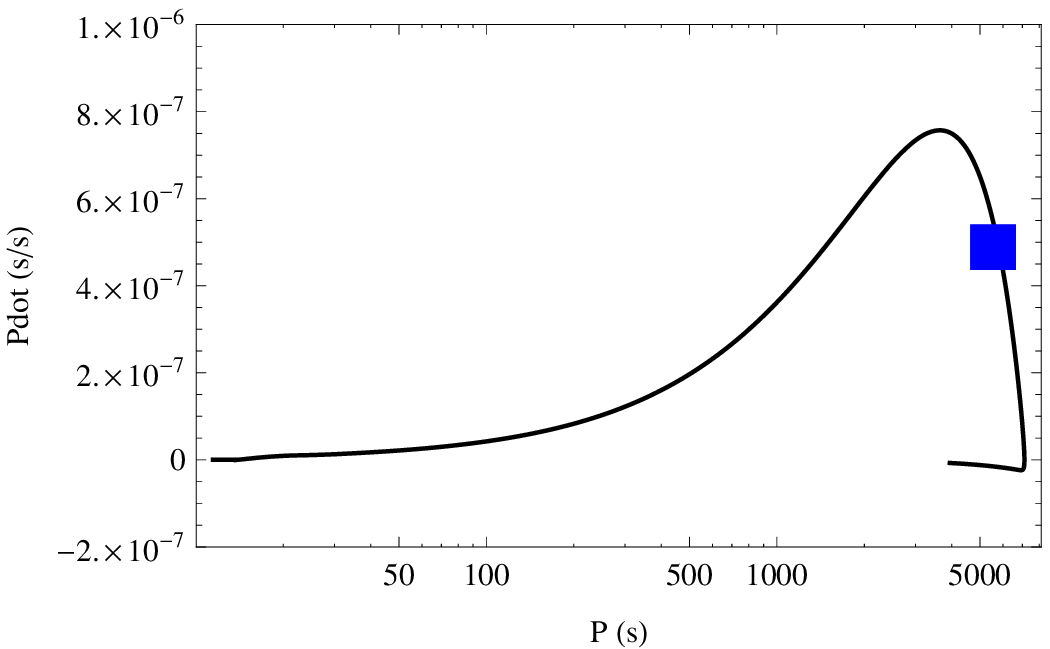}
\end{minipage}
\caption{The rotational evolution of 4U 2206+54. The upper panel is the period evolution with time. It is similar to the dashed line in Figure \ref{fig_P_evo}, except for $B_0=1.6\times 10^{15} \ \rm G$ and $\tau_{\rm H} =5\times 10^3 \ \rm yr$. The middle panel is the corresponding period derivative as a function time. The period derivative is only significant during spin-down epoch. During the ejector phase, the period derivative is very small. During the accretion phase, the corresponding period derivative is about $-1\times 10^{-7}$, as can be seen in the lower panel. The lower panel is the corresponding period derivative as a function of period. In all panels, the black lines and points are the model calculations. The blue dashed lines and blue square are the observations of 4U 2206+54.}
\label{fig_P4U2206}
\end{figure}

In summary, the above two scenarios will both have the torque form of equation (\ref{eqn_Npropeller_2}).  The calculations for 4U 2206+54 is shown in Figure \ref{fig_P4U2206}. The model parameters are similar to the dashed line in Figure \ref{fig_P_evo}, except for a high initial magnetic field of
$B_0=1.6\times 10^{15} \ \rm G$ and a longer Hall time scale of $\tau_{\rm H} =5\times 10^3 \ \rm yr$. A longer Hall time scale ensures that the field has not decayed significantly when the spin-down phase begins. Analytical estimations confirm the numerical results here. Similar calculations for SXP 1062 can also been done which is also spinning down (see previous work by Popov \& Turolla 2012; Fu \& Li 2012). The long period X-ray pulsar SXP 1323 is similar to SXP 1062 (Gvaramadze et al. 2019). However, it showed a possible transition between spin-down and spin-up (Carpano et al. 2017). The spin-up of SXP 1323 has a typical time scale of $\sim 50$ years (Carpano et al. 2017). The spin-up may be due to the formation of transient disks (see the following example for 2S 0114+65).

\subsection{Explanation of the spin-up of 2S 0114+65}

The magnetic field decay during the later phase will result in a decreasing equilibrium period. From the above calculations for
4U 2206+54, the period derivative contributed by the magnetic field decay is about $-1\times 10^{-7} \ \rm s \ s^{-1}$.
The long term spin-up rate of 2S 0114+65 is about $-1\times 10^{-6} \ \rm s \ s^{-1}$ (Hall et al. 2000; Bonning \& Falanga et al. 2005; Farrel et al. 2008; Wang 2011; Sanjurjo-Ferrrin et al. 2017). The analysis of Hu et al. (2017) showed that the long term averaged spin-up rate of 2S 0114+65 is about $-5\times 10^{-7} \ \rm s \ s^{-1}$. So the spin-up rate due to magnetic field decay is one order of magnitude smaller than the observed value in the case of 2S 0114+65. Similar to the NGC 300 ULX1 pulsar
(Tong \& Wang 2019; Vasilopoulos et al. 2018), it is possible that 2S 0114+65 is now running from a previous equilibrium
state to a new equilibrium state. The process may last for hundreds of years (Vasilopoulos et al. 2018). During this process, the neutron star 2S 0114+65 will have a high spin-up rate.

\begin{figure}
\begin{minipage}{0.45\textwidth}
 \includegraphics[width=0.95\textwidth]{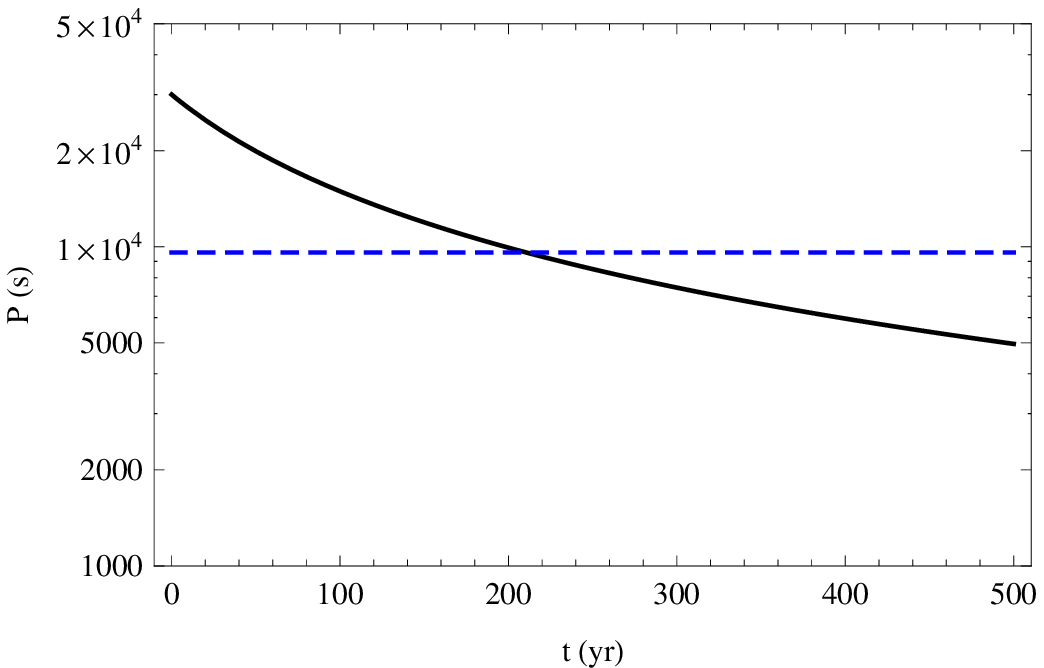}
\end{minipage}
\begin{minipage}{0.45\textwidth}
 \includegraphics[width=0.95\textwidth]{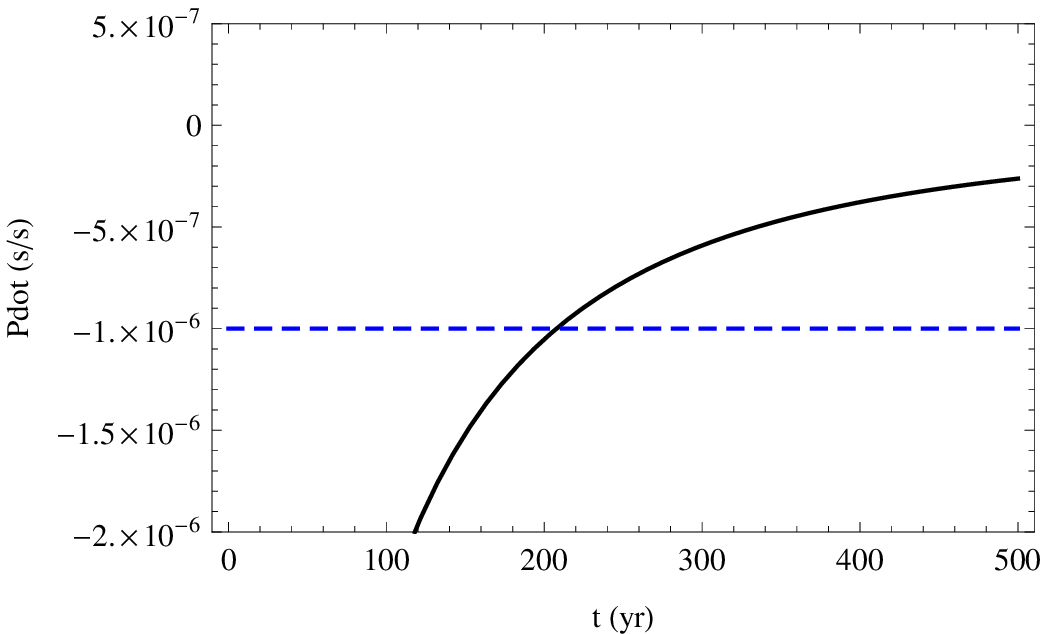}
\end{minipage}
\begin{minipage}{0.45\textwidth}
 \includegraphics[width=0.95\textwidth]{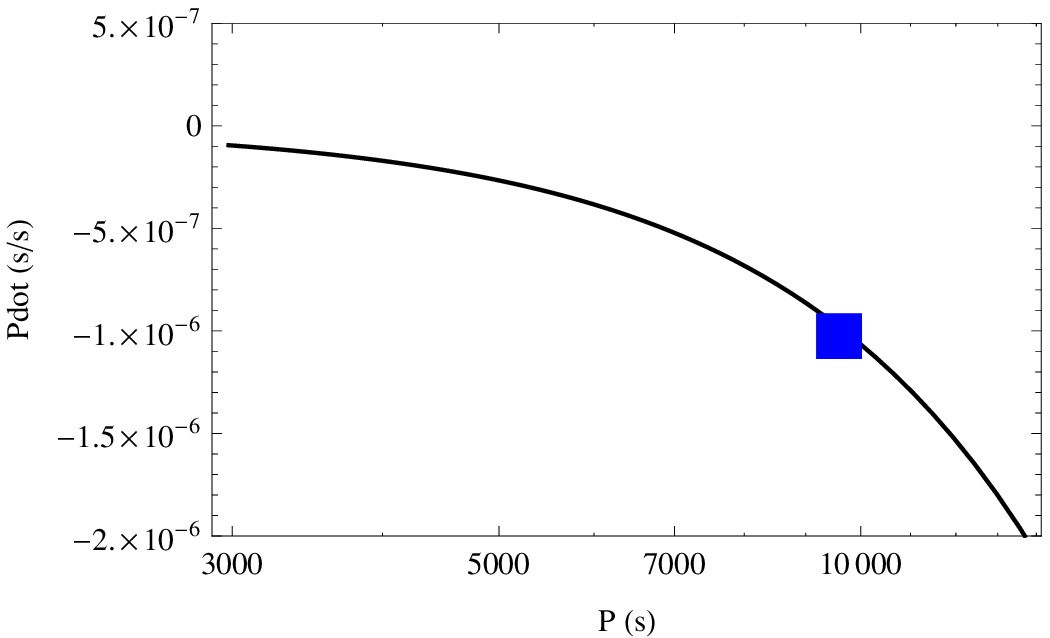}
\end{minipage}
\caption{Short term rotational evolution of 2S 0114+65. The black lines are the model calculations,
the blue dashed lines and blue square are the observations of 2S 0114+65. A starting period of $3\times 10^4 \ \rm s$ is chosen. It may be
reached during the previous equilibrium state. During the short and rapid spin-up phase, the magnetic field is chosen as $10^{12} \ \rm G$, the accretion efficiency parameter is chosen as $\zeta =0.08$, the mass accretion rate is chosen as $2\times 10^{15} \ \rm g \ s^{-1}$. The torque form and magnetic field evolution are the same with those in Figure \ref{fig_P4U2206} and Figure \ref{fig_P_evo}.}
\label{fig_P0114}
\end{figure}

The long term evolution of 2S 0114+65 may be similar to that proposed by Li \& van den Huevel (1999). The neutron star may be born as a magnetar with magnetic field about $10^{15} \ \rm G$. During the main sequence stage of the companion star, the neutron star will spin down to a very long rotational period (e.g., several times of $10^4 \ \rm s$) under a mass accrete rate about $10^{14} \ \rm g \ s^{-1}$. When the companion star enters into supergiant phase, the mass accretion rate may be increase to about $10^{15}-10^{16} \ \rm g \ s^{-1}$. At the same time, the neutron star magnetic field has also decayed significantly, to about $10^{12} \ \rm G$. Due to the nature of wind accretion, the spin period of a neutron star is hardly changed from its previous equilibrium value (Li \& van den Heuvel 1999). Its long term evolution may be determined by the magnetic field decay. However, transient disks may be formed during the wind accretion (Hu et al. 2017). This transient disk may result in the spin-up of the neutron star like that in Figure \ref{fig_P0114}. Analytical calculation of the spin-up rate may be done similar to the disk dominated case (Bildsten et al. 1997; Tong \& Wang 2019). For 2S 0114+65, considering the observed long spin period alone, it seems that a magnetic field of $10^{15} \ \rm G$ and a mass accretion rate of $10^{15} \ \rm g \ s^{-1}$ may be more likely (e.g., from Equation (\ref{eqn_Peq})). But these physical parameters would result in the period derivative which is an order of magnitude higher than the observations. Therefore, by modeling the period and period derivative simultaneously, a magnetic field of $10^{12} \ \rm G$ and mass accretion rate of $10^{15} \rm \ g \ s^{-1}$ are preferred. There are model degenerates within the parameters of magnetic field ($B$), mass accretion rate ($\dot{M}$), and efficiency $\zeta$ in Equation (\ref{eqn_Npropeller_2}). Our estimations offer a quantitative example.

In Figure \ref{fig_P0114}, the initial period is chosen as $3\times 10^4 \ \rm s$, similar to that of AX J1910.7+0917 (Sidoli et al. 2017). If the initial period is chosen as approximately the current period of 2S 0114+65, the evolution timescale can be as short as tens of years, as that observed in 2S 0114+65 (Sanjurjo-Ferrrin et al. 2017; Hu et al. 2017). The calculation here is done for hundreds of years. It can be seen that under the disk torque the period decreases quickly with time. Given enough time, it will approach the equilibrium period given in Equation (\ref{eqn_Peq}). For a magnetic field of $10^{12} \ \rm G$, and mass accretion rate $2\times10^{15} \ \rm g \ s^{-1}$, the equilibrium is about $10$ seconds. However, the transient disk may only last for relatively short time. The transient disk formed at different epoches may switch between the prograde accretion and retrograde accretion. Then the pulsar's spin period will change very little in the long run. The observed spin-up rate of 2S 0114+65 is the long term averaged value. On shorter time scales, it has showed both spin-up and spin-down (Hu et al. 2017). Therefore, a duty cycle $\zeta$ for the transient disk is introduced. For a different $\zeta$, a different mass accretion rate can be chosen in order to explain the observed period and period derivative. At present, 2S 0114+65 showed the spin-up behaviour. It is possible that in the future it may switch to a spin-down state, similar to other accreting X-ray pulsars (Bildsten et al. 1997; Chakrabarty et al. 1997; Gonzalez-Galan et al. 2012).

\section{Discussion and conclusion}

\begin{figure}
  \includegraphics[width=0.55\textwidth]{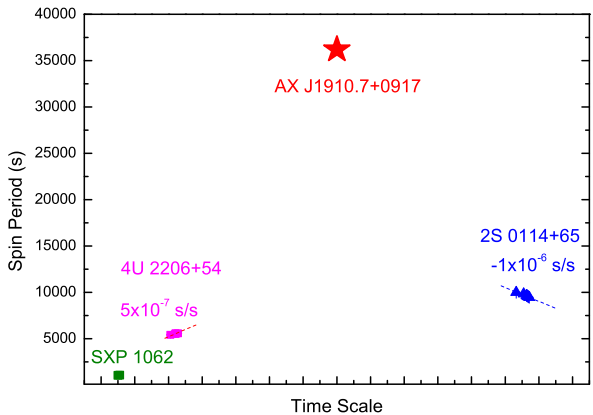}
  \caption{The imaginary picture shows the rotational evolution
of four slow pulsation neutron stars in high-mass X-ray binaries:
SXP 1062, 4U 2206+54, 2S 0114+65 and AX J1910.7+0917.
The dashed lines show their possible spin
evolution channels. 4U 2206+54 undergoes a spin-down process with a
rate of $\sim 5 \times 10^{-7}$ s s$^{-1}$ (from Wang 2013), which will make the spin period of 4U 2206+54 longer than 10 000 s
within 300 years if the rate is stable in the next few hundred years. However,
2S 0114+65 undergoes a fast spin-up process in the last 30 years with
spin-period variations from 10 000 s to present 9300 s (Sanjurjo-Ferrrin et al. 2017). It is
possible that there exists a candidate neutron star binary with a spin period
much longer than 10 000 s which may be the product of the long-term spin
down of the 4U 2206+54-like neutron star and would undergo the spin-up
transition to form the 2S 0114+65-like source. The newly discovered slowest pulsation X-ray pulsar AX J1910.7+0917 in the high mass binary has a spin period of $\sim 36200$ s (the red star, from Sidoli et al. 2017) could be the candidate around the spin period evolution transition phase. }
  \label{fig_slowpsr}
\end{figure}

The long term rotational behaviors of wind accreting magnetars are studied in this work (see Figure \ref{fig_P_evo}).
In modeling the rotational evolution of wind accreting magnetars, there are many uncertainties: magnetic field decay, wind accretion torques, transient disks etc. By considering different modeling of magnetic field decay and accretion torque, it is shown that the coexistence of long term spin-down and spin-up can exist. However, when applying this general picture to the cases of 4U 2206+54 and 2S 0114+65, there are discrepancies between this general modeling and observations of individual sources. Additional inputs are needed, e.g. accretion while spin-down due to outflow or magnetospheric coupling for 4U 2206+54, transient disks in 2S 0114+65. Despite these uncertainties and additional astrophysical inputs, we want to show that the origin of long spin period, coexistence of spin-down and spin-up of long period X-ray pulsars are still understandable in the wind accreting magnetar scenario.

For the general case of accreting neutron stars, several sources showed abrupt changes between spin-up and spin-down (Bildsten et al. 1997; Chakrabarty et al. 1997; Gonzalez-Galan et al. 2012). The absolute torques in the spin-up and spin-down case are roughly the same. The typical timescale can be tens of years. At the same time, Be/X-ray pulsars in the small Magellanic clouds contain both spin-up and spin-down sources (Yang et al. 2017; Christodoulou et al. 2017). The torque reversal also happens on a time scale of tens of years (Christodoulou et al. 2017). These spin-up/spin-down behaviors may be caused by the transition between prograde and retrograde accretion disks (Chakrabarty et al. 1997; Christodoulou et al. 2017). The long period X-ray pulsar 2S 0114+65 also showed random spin-up and spin-down on even shorter time scale ($\sim$ 1 yr., Hu et al. 2017). It may be caused by the random angular momentum during wind accretion. While the spin-up of 2S 0114+65 (which can lasts for tens of years) may be due to the presence of transient accretion disks (Hu et al. 2017). In different time scales, the spin-up/spin-down behavior may be governed by different physical processes. These processes may happen simultaneously in the same source.

Observationally, 4U 2206+54 has a main sequence companion (Blay et al. 2006), while 2S 0114+65's companion star is a supergiant (Reig et al. 1996).
This is also consistent with our general expectations (Figure \ref{fig_P_evo}): young systems will spin down while older systems will spin up.
The wind accreting magnetar will evolve from 4U 2206+54-like sources to 2S 0114+65-like sources. Then some missing link sources with very long pulsation period between 4U 2206+54 and 2S 0114+65 may be expected (Wang 2013). The recently confirmed long period X-ray pulsars AX J1910.7+0917 has
a rotational period about 10 hours (Sidoli et al. 2017). We think that AX J1910.7+0917 may be the missing link source between 4U 2206+54 and 2S 0114+65.
Possible evolutional scenario between these sources are presented in Figure \ref{fig_slowpsr}. Future monitoring observations on AX J1910.7+0917 by X-ray missions are needed to derive the possible spin evolution in the long run. It is possible that the spin-down or spin-up rate of AX J1910.7+0917 lies between that of 4U 2206+54 or 2S 0114+65, from $\dot{P} \sim -10^{-6}\ \rm s \ s^{-1}$ to $10^{-6} \ \rm s \ s^{-1}$. More samples will help to understand or confirm the evolutional channel of this special class of X-ray pulsars as shown in Figure \ref{fig_slowpsr}.

There may be transient disks in wind accreting magnetars. Then the long term rotational evolution of the neutron star may be dominated by the disk torque due to the effectiveness of disk accretion. Then the $\zeta$ parameter in equation (\ref{eqn_Npropeller_2}) will reflect the duty cycle of the transient accretion disk. Possible evidences include: (1) The long term rotational behavior of 2S 0114+65 may hint the existence of a transient accretion disk (Hu et al. 2017). Possible evidence for a transient disk in other super-giant high mass X-ray binaries were also found (Koh et al. 1997; Jenke et al. 2012; Taani et al. 2019). (2) 4U 2206+54 may be an atypical Be X-ray binary (Reig et al. 2011). Transient disks in typical Be X-ray binaries are ubiquitous (Bildsten 1997; Wilson et al. 2008; Sugizaki et al. 2017). (3) The reason for the formation of transient disks in wind accreting neutron star may be due to a low wind velocity (Frank et al. 2002) or a closer system (EI Mallah et al. 2019). 2S 0114+65 and 4U 2206+54 have the smallest orbital period compared with other long period X-ray pulsars (see Figure 4 in Tong \& Wang 2019). The stellar wind velocity in the case of 4U 2206+54 is very low (Reig et al. 2012). It will also make the formation of disks easier.

In conclusion, the period and period derivative of long period X-ray pulsars should be modeled simultaneously. The coexistence of both spin-up and spin-down behaviors should also be understandable in the same theoretical frame. We have tried to model the rotational behaviors of long period X-ray pulsars from the wind accreting magnetar point of view, considering the effect of magnetic field decay. Despite the several uncertainties, wind accreting magnetars may be an alternative model for long period X-ray pulsars.

\section*{Acknowledgements}

We are very grateful to the referee for fruitful suggestions to significantly improve the manuscript.
W. Wang is supported the National Program on Key Research and Development Project (Grants No. 2016YFA0400803) and the NSFC (11622326 and U1838103). H. Tong is supported by the NSFC (11773008).

\appendix
\section{Different spin-up and spin-down torques}

The discussions here are inspired by Ikhsanov (2007) and Fu \& Li (2012), and is similar to that in Tong et al. (2016). The rotational evolution of an accreting neutron star
is governed by (Ikhsanov 2007)
\begin{equation}
I \frac{{\rm d}\Omega}{{\rm d}t} = N_{\rm su}+N_{\rm sd},
\end{equation}
where total torque is split into spin-up part ($N_{\rm su}$) and spin-down part ($N_{\rm sd}$). When the two balances, an accretion equilibrium period is reached. For the case of disk accretion, the spin-up torque is
\begin{equation}
N_{\rm su}^{\rm d}  = \dot{M}_{\rm acc} R_{\rm m}^2 \Omega_{\rm K}(R_{\rm m}) = \dot{M}_{\rm acc} \sqrt{G M_{\rm ns} R_{\rm m}} .
\end{equation}
In the case of wind accretion, the spin-up torque is (Frank et al. 2002)
\begin{equation}
N_{\rm su}^{\rm w} = \dot{M}_{\rm acc} \times \frac14 \Omega_{\rm orb} R_{\rm acc}^2,
\end{equation}
where $\Omega_{\rm orb} =2\pi/P_{\rm orb}$ is the orbital angular velocity, $P_{\rm orb}$ is orbital period of the neutron binary system. The spin-up torque will be suppressed in the wind accretion case compared with that of the disk accretion case
\begin{eqnarray}
\zeta &=&\frac{N_{\rm su}^{\rm w}}{N_{\rm su}^{\rm d}} \\
&=& 0.026 \dot{M}_{15}^{1/7} \mu_{33}^{-2/7} \left( \frac{P_{\rm orb}}{19.12 \ \rm day}\right)^{-1}
\left( \frac{V}{350 \ \rm km \ s^{-1}} \right)^{-4},
\end{eqnarray}
where typical orbital period and wind velocity of 4U 2206+54 are inserted (Reig et al. 2012; Wang 2013). This value is consistent with the typical values used in equation (\ref{eqn_Npropeller_2}).

There are many uncertainties in the spin-down torque. A commonly used maximum spin-down torque is (Ikhsanov 2007; Fu \& Li 2012 and references therein)
\begin{equation}\label{eqn_Nspindownmax}
N_{\rm sd,max} = - \frac{\mu^2}{R_{\rm co}^3}.
\end{equation}
Considering the possibility of energy budget and angular budget, the spin-down torque may be expressed in a power law form
(Mori \& Ruderman 2003; Fu \& Li 2012)
\begin{equation}\label{eqn_Nspindowngamma}
N_{\rm sd, \gamma} = - \dot{M}_{\rm acc} R_{\rm m}^2 \Omega_{\rm K}(R_{\rm m})
\left ( \frac{\Omega}{\Omega_{\rm K}(R_{\rm m})}  \right)^{\gamma},
\end{equation}
where the dimensionless parameter $\gamma$ may have values of  $-1, 0, 1, 2$ etc. Comparion between equation (\ref{eqn_Nspindownmax}) and equation (\ref{eqn_Nspindowngamma}) shows that equation (\ref{eqn_Nspindownmax}) corresponds to the case of $\gamma=2$ (although the normalization is different). If both the spin-up and spin-down torque are suppressed in the wind accretion case, then the total accretion torque  will be
\begin{equation}\label{eqn_Ntot_gamma}
N_{\rm tot} = \zeta (N_{\rm su} + N_{\rm sd}) =  \zeta \dot{M}_{\rm acc} R_{\rm m}^2 \Omega_{\rm K}(R_{\rm m}) \left[ 1- \left( \frac{\Omega}{\Omega_{\rm K}(R_{\rm m})} \right)^{\gamma}  \right].
\end{equation}
Comparison between equation (\ref{eqn_Npropeller_2}) and equation (\ref{eqn_Ntot_gamma}) shows that the calculations in the main text is for the case of $\gamma=1$. Similar calculations for $\gamma=-1$ shows that it is very inefficient in spinning down the neutron star (consistent with the conclusions of Fu \& Li 2012). For $\gamma=2$ or using the maximum spin-down torque, the result is similar to the case of $\gamma=1$ (i.e. result in figure \ref{fig_P_evo}). Irrespective of the exact value of the $\gamma$ parameter, the
equilibrium period will always be the same (as can be seen in equation (\ref{eqn_Ntot_gamma})). Combined with idea of magnetic field decay in the long term, the wind accreting magnetar will always spin-down in the early stage (to catch the equilibrium period) and spin-up in the later stage (to a decreasing equilibrium period).

If only the spin-up torque is suppressed in the wind accreting case, this will mainly result in a longer equilibrium period (Ikhsanov 2007).
If a transient disk is formed in the wind accretion case, then the long term rotational evolution of the neutron star may be dominated by the disk torque. Then the total accretion torque will be similar to equation (\ref{eqn_Npropeller_2}) (or equation \ref{eqn_Ntot_gamma}), except that the $\zeta$ factor reflects the duty cycle of the transient disk. This is the reason we prefer to use equation (\ref{eqn_Npropeller_2}) (or equation \ref{eqn_Ntot_gamma}).







\bsp	
\label{lastpage}
\end{document}